\input{psfig}
\input{epsf}

\documentstyle[12pt,lathuile]{article}
\def\simge{\mathrel{%
   \rlap{\raise 0.511ex \hbox{$>$}}{\lower 0.511ex \hbox{$\sim$}}}}
\def\simle{\mathrel{
   \rlap{\raise 0.511ex \hbox{$<$}}{\lower 0.511ex \hbox{$\sim$}}}}
 
\def\slashchar#1{\setbox0=\hbox{$#1$}           
   \dimen0=\wd0                                 
   \setbox1=\hbox{/} \dimen1=\wd1               
   \ifdim\dimen0>\dimen1                        
      \rlap{\hbox to \dimen0{\hfil/\hfil}}      
      #1                                        
   \else                                        
      \rlap{\hbox to \dimen1{\hfil$#1$\hfil}}   
      /                                         
   \fi}                                         %

\def\thc{\theta_C}
\def\thy{\theta_Y}
\def\nn{\nonumber}
\def\ts{\thinspace}

\def\ra{\rightarrow}

\def\ol{\bar}

\def\be{\begin{equation}} 
\def\ee{\end{equation}} 
\def\bea{\begin{eqnarray}}
\def\eea{\end{eqnarray}}
\def\ba{\begin{array}}
\def\ea{\end{array}}

\def\CD{{\cal D}}

\def\CH{{\cal H}}

\def\CM{{\cal M}}

\def\CO{{\cal O}}

\def\CU{{\cal U}}

\def\chtct{\CH_{TC2}}

\def\chetc{\CH_{ETC}}

\def\atc{\alpha_{TC}}

\def\Ntc{N_{TC}}

\def\sutc{SU(\Ntc)}

\def\getc{g_{ETC}}

\def\uone{U(1)_1}
\def\utwo{U(1)_2}
\def\uy{U(1)_Y}
\def\suone{SU(3)_1}
\def\sutwo{SU(3)_2}

\def\kslash{\raise.15ex\hbox{/}\kern-.57em k}
\def\LTC{\Lambda_{TC}}

\def\METC{M_{ETC}}

\def\Mv{M_{V_8}}
\def\Mzp{M_{Z'}}

\def\condtc{{\langle \ol T T \rangle}_{TC}}
\def\condetc{{\langle \ol T T \rangle}_{ETC}}

\def\mev{{\rm MeV}}
\def\gev{{\rm GeV}}
\def\tev{{\rm TeV}}

\def\half{{\textstyle{ { 1\over { 2 } }}}}

\def\nin{\noindent}
\makeatletter
\def\@cite#1#2{[{#1\if@tempswa , #2\fi}]}
\makeatother
\begin{document}
\title{ 
$K^0$--$\bar K^0$ and $B^0$--$\bar B^0$ CONSTRAINTS ON TECHNICOLOR
}
\author{KENNETH LANE
\\
{\em Department of Physics, Boston University}\\
{\em 590 Commonwealth Avenue, Boston, Massachusetts 02215} \\
}
\maketitle
\baselineskip=14.5pt
\begin{abstract}
I propose a resolution of the strong CP problem of quarks based on vacuum
alignment in theories of dynamically broken electroweak and flavor
symmetries. Mixing in the neutral $B$ and $K$--meson systems is used to
restrict the form of quark mass and mixing matrices and to constrain the
masses of the topcolor gauge bosons $V_8$ and $Z'$ to be greater than
5~TeV. The $K^0$ mixing parameter $\epsilon$ is calculated for models whose
quark mass matrix leads to weak, but not strong, CP violation. It is easy to
accomodate its value in some models while, in others, no reasonable
combination of extended technicolor and topcolor masses can account for
$\epsilon$.
\end{abstract}
\baselineskip=17pt
\newpage
\section*{1. Outline}

In this talk I discuss the dynamical approach to CP violation in technicolor
theories and a few of its consequences. Special attention will be paid to
constraints from the $K^0$ and $B_d^0$ systems. I will cover the following
topics:~\footnote{Invited talk at Les Rencontres de Physique de la Vallee
d'Aoste, La Thuile, Italy, March 4-10, 2001. Most of the work reported here
was done in collaboration with Gustavo Burdman, Estia Eichten and Tongu\c c
Rador.}

\begin{enumerate}

\item{} Vacuum alignment in technicolor theories and the rational--phase
  solutions.

\item{} A proposal to solve the strong CP problem without an axion or a
  massless up quark.

\item{} The structure of quark mass and mixing matrices in extended
technicolor (ETC) theories with topcolor--assisted technicolor (TC2). In
particular, realistic Cabbibo--Kobayashi--Maskawa (CKM) matrices are easily
generated.

\item{} Flavor--changing neutral current interactions from extended
technicolor and topcolor.

\item{} Strong constraints on TC2 gauge boson masses from $B_d$--$\ol B_d$
  mixing.

\item{} New results on the $K^0$--$\ol K^0$ CP--violating parameter
$\epsilon$ and constraints on ETC masses.

\end{enumerate}

\section*{2. Vacuum Alignment in the Technifermion Sector}

In 1971, Dashen stressed the importance of matching the ground state
$|\Omega\rangle$ of a theory containing spontaneously broken chiral
symmetries with the perturbing Hamiltonian $\CH'$ that explicitly breaks
those symmetries~\cite{rfd}. He also showed that this process, known as vacuum
alignment, can lead to a spontaneous breakdown of CP invariance: it may
happen that the CP symmetry of $|\Omega\rangle$ is not the same as that of
the the aligned $\CH'$. This idea found its natural home in dynamical
theories of electroweak symmetry breaking---technicolor~\cite{tc}---because
they have large groups of flavor/chiral symmetries that are spontaneously
broken by strong dynamics and explicitly broken by extended technicolor
(ETC)~\cite{etc,tcreview,rscreview}. Furthermore, the perturbation $\CH'$
generated by exchange of ETC gauge bosons is {\it naively} CP--conserving if
CP is unbroken above the technicolor energy scale. Thus, in 1979, Eichten,
Preskill and I proposed that CP violation occurs spontaneously in theories of
dynamical electroweak symmetry breaking~\cite{align}. Our goal, unrealized at
the time, was to solve the strong CP problem of QCD {\it without} invoking a
Peccei--Quinn symmetry or a massless up quark~\cite{CPreview}.

This problem was taken up again a few years ago with Eichten and
Rador~\cite{vacalign}. We studied the first important step in reaching this
goal: vacuum alignment in the technifermion sector. We considered models in
which a single kind of technifermion interacts with quarks via ETC
interactions. There are $N$ technifermion doublets $T_{L,R\ts I} =
(\CU_{L,R\ts I}, \ts \CD_{L,R\ts I})$, $I = 1,2,\dots,N$, all transforming
according to the fundamental representation of the technicolor gauge group
$SU(N_{TC})$. There are 3~generations of $SU(3)_C$ triplet quarks $q_{L,R\ts
i} = (u_{L,R\ts i}, \ts d_{L,R\ts i})$, $i = 1,2,3$. The left--handed
fermions are electroweak $SU(2)$ doublets and the right--handed ones are
singlets. Here and below, we exhibit only flavor, not technicolor nor color,
indices.

The technifermions are assumed for simplicity to be ordinary color--singlets,
so the chiral flavor group of our model is $G_f = \left[SU(2N)_{L} \otimes
SU(2N)_{R}\right] \otimes\left[SU(6)_{L} \otimes
SU(6)_{R}\right]$.~\footnote{The fact that heavy quark chiral symmetries
cannot be treated by chiral perturbative methods will be addressed below. We
have excluded anomalous $U_A(1)$'s strongly broken by TC and color instanton
effects.} When the TC and QCD couplings reach their required critical values,
these symmetries are spontaneously broken to $S_f = SU(2N) \otimes SU(6)$. We
adopt as a ``standard vacuum'' on which to carry out chiral perturbation
theory the ground state $|\Omega\rangle$ whose symmetry group is the the
vectorial $SU(2N)_V \otimes SU(6)_V$, with fermion bilinear condensates given
by
\bea\label{eq:standard}
\langle \Omega |\ol \CU_{LI} \CU_{RJ}|\Omega \rangle &=&
\langle \Omega |\ol \CD_{LI} \CD_{RJ}|\Omega \rangle = -\delta_{IJ} \Delta_T
\nn\\
\langle \Omega |\ol u_{Li} u_{Rj}|\Omega \rangle &=&
\langle \Omega |\ol d_{Li} d_{Rj}|\Omega \rangle = -\delta_{ij} \Delta_q \ts.
\eea
Here, $\Delta_T \simeq N_{TC} \Lambda^3_{TC}$ and $\Delta_q \simeq
N_C \Lambda^3_{QCD}$ when they are renormalized at their respective strong
interaction scales. Of course, $N_C = 3$.

All of the $G_f$ symmetries except for the gauged electroweak $SU(2) \otimes
U(1)$ must be {\it explicitly} broken by extended technicolor
interactions~\cite{etc}. In the absence of a concrete ETC model, we write the
interactions broken at the scale $\METC/\getc \sim
10^2$--$10^4\,\tev$~\footnote{See the Appendix for estimates of
$\METC/\getc$.} in the phenomenological four-fermion form (sum over
repeated indices)~\footnote{We assume that ETC interactions commute with
  electroweak $SU(2)$, though not with $U(1)$ nor color $SU(3)$. All fields
  in Eq.~(2) are electroweak, not mass, eigenstates.}
\bea\label{eq:Hetc}
\CH' &\equiv& \CH'_{TT} + \CH'_{Tq} + \CH'_{qq} \nn\\
&=& \Lambda^{TT}_{IJKL} \ts \ol{T}_{LI}\gamma^{\mu}T_{LJ}
\ts \ol{T}_{RK}\gamma_{\mu}T_{RL} \nn 
+ \Lambda^{Tq}_{IijJ} \ts \ol{T}_{LI}\gamma^{\mu}q_{Li}
\ts \ol{q}_{Rj}\gamma_{\mu}T_{RJ} + {\rm h.c.}\\
&+& \Lambda^{qq}_{ijkl} \ts \ol{q}_{Li}\gamma_{\mu}q_{Lj}
\ts \ol{q}_{Rk}\gamma_{\mu}q_{Rl} \ts,
\eea
where $T_{L,R\ts I}$ and $q_{L,R\ts i}$ stand for all $2N$ technifermions and
6~quarks, respectively. Here, $\METC$ is a typical ETC gauge boson mass and
the $\Lambda$ coefficients are $\CO(g^2_{ETC}/M^2_{ETC})$ times mixing
factors for these bosons and group theoretical factors. The $\Lambda$'s may
have either sign. In all calculations, we must choose the $\Lambda$'s to
avoid unwanted Goldstone bosons. Hermiticity of $\CH'$ requires
\be\label{eq:herm}
(\Lambda^{TT}_{IJKL})^* = \Lambda^{TT}_{JILK} \ts, \qquad
(\Lambda^{Tq}_{IijJ})^* = \Lambda^{Tq}_{iIJj} \ts, \qquad
(\Lambda^{qq}_{ijkl})^* = \Lambda^{qq}_{jilk} \ts.
\ee
The assumption of time--reversal invariance for this theory before any
potential breaking via vacuum alignment means that the instanton angles
$\theta_{TC} = \theta_{QCD} = 0$ (at tree level) and that all $\Lambda$'s are
real. Thus, e.g., $\Lambda^{TT}_{IJKL} = \Lambda^{TT}_{JILK}$.

Having chosen a standard chiral--perturbative ground state, $|\Omega\rangle$,
vacuum alignment proceeds by minimizing the expectation value of the rotated
Hamiltonian. This is obtained by making the $G_f$ transformation $T_{L,R} \ra
W_{L,R} \ts T_{L,R}$ and $q_{L,R} \ra Q_{L,R} \ts q_{L,R}$, where $W_{L,R}
\in SU(2N)_{L,R}$ and $Q_{L,R} \in SU(6)_{L,R}$:~\footnote{So long as vacuum
  alignment preserves electric charge conservation, the alignment matrices
  will be block--diagonal:
\bea\label{block}
W_{L,R} =  \left(\ba{cc} W^U & 0 \\ 0 & W^D \ea\right)_{L,R} \ts, \qquad
Q_{L,R} =  \left(\ba{cc} U & 0 \\ 0 & D \ea\right)_{L,R} \ts. \nn
\eea}
\bea\label{eq:HW}
\CH'(W,Q) &=& \CH'_{TT}(W_L,W_R) +  \CH'_{Tq}(W,Q) +
\CH'_{qq}(Q_L,Q_R) \\
&=& \Lambda^{TT}_{IJKL} \ts \ol{T}_{LI'} W_{L\ts I'I}^\dag
\gamma^{\mu}W_{L\ts JJ'}T_{LJ'} \ts \ol{T}_{RK'} W_{R\ts K'K}^\dag
\gamma^{\mu}W_{R\ts LL'}T_{RL'} + \cdots \ts.\nn
\eea
Since $T$ and $q$ transform according to complex representations of their
respective color groups, the four--fermion condensates in the
$S_f$--invariant $|\Omega\rangle$ have the form
\bea\label{eq:conds}
\langle\Omega|\ol{T}_{LI}\gamma^{\mu}T_{LJ}
\ts \ol{T}_{RK}\gamma_{\mu}T_{RL}|\Omega\rangle &=& -\Delta_{TT}
\delta_{IL}\delta_{JK} \ts, \nonumber \\
\langle\Omega| \ol{T}_{LI}\gamma^{\mu}q_{Li}
\ts \ol{q}_{Rj}\gamma_{\mu}T_{RJ} |\Omega\rangle &=&
-\Delta_{Tq}\delta_{IJ}\delta_{ij} \ts, \\
\langle\Omega|\ol{q}_{Li}\gamma^{\mu}q_{Lj}
\ts \ol{q}_{Rk}\gamma_{\mu}q_{Rl} |\Omega\rangle &=&
-\Delta_{qq}\delta_{il}\delta_{jk} \ts. \nonumber 
\eea
The condensates are positive, renormalized at $\METC$ and, in the
large--$N_{TC}$ and $N_C$ limits, they are given by $\Delta_{TT} \simeq
(\Delta_T(\METC))^2$, $\Delta_{Tq} \simeq \Delta_T(\METC)
\Delta_q(\METC)$, and $\Delta_{qq} \simeq (\Delta_q(\METC))^2$. In
walking technicolor~\cite{wtc}, $\Delta_T(\METC) \simeq
(\METC/\Lambda_{TC}) \ts \Delta_T(\Lambda_{TC})$ $= 10^2$--$10^4 \times
\Delta_T(\Lambda_{TC})$. In QCD, however, $\Delta_q(\METC) \simeq
(\log(\METC/\Lambda_{QCD}))^{\gamma_m} \ts \Delta_q(\Lambda_{QCD}) \simeq
\Delta_q(\Lambda_{QCD})$, where $\gamma_m \simeq 2\alpha_C/\pi$ for
$SU(3)_C$. Thus, the ratio
\be\label{eq:ratio}
r = {\Delta_{Tq}(\METC) \over{\Delta_{TT}(\METC)}} \simeq
{\Delta_{qq}(\METC) \over{\Delta_{Tq}(\METC)}}
\ee
is at most $10^{-10}$. This is 10 to $10^4$ times smaller than in a
technicolor theory in which the coupling does not walk.

With these condensates, the vacuum energy is a function only of $W = W_L \ts
W_R^\dag$ and $Q = Q_L \ts Q_R^\dag$, elements of the coset space $G_f/S_f$:
\bea\label{eq:vacE}
& &E(W,Q) = E_{TT}(W) + E_{Tq}(W,Q) + E_{qq}(Q) \\
& & \ts\ts = -\Lambda^{TT}_{IJKL} \ts W_{JK} \ts W^\dag_{LI} \ts \Delta_{TT}
       -\left(\Lambda^{Tq}_{IijJ} \ts Q_{ij} \ts W^\dag_{JI} + {\rm c.c.}
         \right) \Delta_{Tq} 
       -\Lambda^{qq}_{ijkl} \ts Q_{jk} \ts Q^\dag_{li} \ts \Delta_{qq} \nn \\
& & \ts\ts = -\Lambda^{TT}_{IJKL} \ts W_{JK} \ts W^\dag_{LI} \ts \Delta_{TT}
+\CO(10^{-10}) \ts.\nn 
\eea
Note that time--reversal invariance of the unrotated Hamiltonian $\CH'$
implies that $E(W,Q) = E(W^*,Q^*)$. Hence, spontaneous CP violation occurs
if the solutions $W_0$, $Q_0$ to the minimization problem are complex.

The last line of Eq.~(\ref{eq:vacE}) makes clear that we should first
minimize the technifermion sector energy $E_{TT}$. This determines $W_0$ up
to corrections of $\CO(10^{-10})$.~\footnote{Two sorts of corrections to this
statement are under study. The first are higher--order ETC and electroweak
corrections to $E_{TT}$. The second are due to $\ol T t \ol t T$ terms in
$E_{Tq}$ which are important if the top condensate is large. I thank
J.~Donoghue and S.~L.~Glashow for emphasizing the importance of these
corrections.} This result is then fed into $E_{Tq}$ to determine $Q_0$---and
the nature of quark CP violation---up to corrections which are also
$\CO(10^{-10})$.

In Ref.~\cite{vacalign} it was shown that just three possibilities naturally
occur for the phases in $W$. (We drop the subscript ``0'' from now on.) Let
us write $W_{IJ} = |W_{IJ}| \exp{(i\phi_{IJ})}$. Consider an individual term
$-\Lambda^{TT}_{IJKL} \ts W_{JK} \ts W^\dag_{LI} \ts \Delta_{TT}$ in the
vacuum energy. If $\Lambda^{TT}_{IJKL} > 0$, this term is least if $\phi_{IL}
= \phi_{JK}$; if $\Lambda^{TT}_{IJKL} < 0$, it is least if $\phi_{IL} =
\phi_{JK} \pm \pi$. We say that $\Lambda^{TT}_{IJKL} \neq 0$ links
$\phi_{IL}$ and $\phi_{JK}$, and tends to align (or antialign) them. Of
course, the constraints of unitarity may partially or wholly frustrate this
alignment. The three possibilities for the phases are:

\begin{enumerate}

\item{} The phases are all unequal, irrational multiples of $\pi$ that are
random except for the constraints of unitarity and unimodularity.

\item{} All of the phases may be equal to the same integer multiple of
$2\pi/N$ (mod~$\pi$). This occurs when all phases are linked and aligned, and
the value $2\pi/N$ is a consequence of unimodularity.~\footnote{Because $W$
is block diagonal, $E_{TT}$ factorizes into two pieces, $E_{UU} + E_{DD}$, in
which $W_U$ and $W_D$ may each be taken unimodular. Thus, totally aligned
phases are multiples of $2\pi/N$, not $\pi/N$.} In this case we say that the
phases are ``rational''.

\item{} Several groups of phases may be linked among themselves and the
phases only partially aligned. In this case, their values are various
rational multiples of $\pi/N'$ for one or more integers $N'$ from~1 to~$N$.

\end{enumerate}

\nin We stress that, as far as we know, rational phases occur naturally only
in ETC theories. They are a consequence of $E_{TT}$ being quadratic, not
linear, in $W$. With these three outcomes in hand, we proceed to investigate
the strong CP violation problem of quarks.

\section*{3. A Dynamical Solution to the Strong CP Problem}

There are two kinds of CP violation in the quark sector. Weak CP violation
enters the standard weak interactions through the CKM phase $\delta_{13}$
and, for us, in the ETC and TC2~\cite{tctwohill} interactions through phases
in the quark alignment alignment matrices $U_{L,R}$ and $D_{L,R}$ discussed
in Section~4. Strong CP violation, which can produce electric dipole moments
$10^{16}$ times larger than in the standard model, is a consequence of
instantons~\cite{CPreview}. No discussion of the origin of CP violation is
complete which does not eliminate strong CP violation. Resolving the strong
CP problem amounts to making $\ol \theta_q = \arg\det(M_q) \simle 10^{-10}$
(in a basis with instanton angle $\theta_{QCD} = 0$), so that the neutron
electric dipole moment is below its experimental bound of $0.63\times
10^{-25}\,e$--cm~\cite{pdg}. Here, $M_q$ is the hard or current algebra mass
matrix of the quarks. It includes both the TC2 and ETC--generated parts of
the top quark's mass.

The ``primordial'' quark mass matrix, the coefficient of the bilinear $\ol
q'_{Ri} q'_{Lj}$ of quark electroweak eigenstates, is generated by ETC
interactions and is given by\footnote{The matrix element $\CM_{tt}$ arises
almost entirely from the TC2--induced condensation of top quarks. We assume
that $\langle \ol t t \rangle$ and $\CM_{tt}$ are real in the basis with
$\theta_{QCD} = 0$. Since technicolor, color, and topcolor groups are
embedded in ETC, all CP--conserving condensates are real in this basis.}
\be\label{eq:primordial}
 (\CM_q)_{ij} = \sum_{I,J} \Lambda^{Tq}_{IijJ} \ts W^\dag_{JI} \ts
 \Delta_T(\METC)  \qquad (q,T = u,U \ts\ts {\rm or} \ts\ts d,D)\ts.
\ee
The $\Lambda^{Tq}_{IijJ}$ are real ETC couplings of order
$(10^2$--$10^4\,\tev)^{-2}$ (see the Appendix). Furthermore, the quark
alignment matrices $Q_{L,R}$ which diagonalize $\CM_q$ to $M_q$ are
unimodular. Thus, $\arg\det(M_q) = \arg\det(\CM_q) \equiv \arg\det(\CM_u) +
\arg\det(\CM_d)$, and the question of strong CP violation is determined {\it
entirely} by the character of vacuum alignment in the technifermion sector,
i.e., by the phases $\phi_{IJ}$ of $W$, and by how the ETC factors
$\Lambda^{Tq}_{IijJ}$ map these phases into the $(\CM_q)_{ij}$.

If the $\phi_{IJ}$ are random irrational phases, $\ol \theta_q$ could vanish
only by the most contrived, unnatural adjustment of the $\Lambda^{Tq}$. If
all $\phi_{IJ} = 2m\pi/N$ (mod $\pi$), then all elements of $\CM_u$ have the
same phase, as do all elements of $\CM_d$. Then, $U_{L,R}$ and $D_{L,R}$ will
be real orthogonal matrices, up to an overall phase. There may be strong
CP violation, but there will no weak CP violation in any interaction.

There remains the possibility, which we assume henceforth, that the
$\phi_{IJ}$ are different rational multiples of $\pi$. Then, strong CP
violation will be absent if the $\Lambda^{Tq}$ map these phases onto the
primordial mass matrix so that each element $(\CM_q)_{ij}$ has a rational
phase {\it and} these add to zero in $\arg\det(\CM_q)$. In the absence of an
explicit ETC model, we are not certain this can happen, but we see no reason
that it cannot. For example, there may be just one nonzero
$\Lambda^{Tq}_{IijJ}$ for each pair $(ij)$ and $(IJ)$. An ETC model which
achieves such a phase mapping will solve the strong CP problem, i.e., $\ol
\theta_q \simle 10^{-10}$, without an axion and without a massless up
quark. This is, in effect, a ``natural fine--tuning'' of phases in the quark
mass matrix.~\footnote{I thank C.~Sommerfield for this description.} There
is, of course, no reason weak CP violation will not occur in this model. We
shall illustrate this with some examples in Sections~4 and~6.

Determining the quark alignment matrices $Q_{L,R}$ begins with minimizing the
vacuum energy
\be\label{eq:EqT}
E_{Tq}(Q) \cong -\half {\rm Tr}\left(\CM_q Q + {\rm
    h.c.}\right)\Delta_q(\METC)
\ee
to find $Q=Q_L Q^\dag_R$. Whether or not $\ol \theta_q = 0$, the matrix
$Q^\dag \CM_q$ is hermitian up to the identity matrix~\cite{rfd},
\be\label{eq:nuyts}
 \CM_q Q - Q^\dag \CM^\dag_q = i\nu_q \ts 1 \ts,
\ee
where $\nu_q$ is the Lagrange multiplier associated with the unimodulariy
constraint on $Q$, and $\nu_q$ vanishes if $\ol \theta_q$ does. Thus, $Q^\dag
\CM_q$ may be diagonalized by the single unitary transformation $Q_R$ and
so~\footnote{Since quark vacuum alignment is based on first order chiral
perturbation theory, it is inapplicable to the heavy quarks $c,b,t$. When
$\ol \theta_q = 0$, Dashen's procedure is equivalent to making the mass
matrix diagonal, real, and positive. Thus, it correctly determines the quark
unitary matrices $U_{L,R}$ and $D_{L,R}$ and the magnitude of strong and weak
CP violation.}
\be\label{eq:Mdiag}
M_q \equiv \left(\ba{cc} M_u & 0 \\ 0 & M_d \ea\right) = Q^\dag_R \CM_q
\ts Q Q_R = Q^\dag_R \CM_q \ts Q_L \ts.
\ee

\section*{4. Quark Mass and Mixing Matrices in ETC/TC2}

\subsection*{4.1 General Considerations}

If $\ol \theta_q = 0$, the matrix $\CM_q$ is brought to real, positive,
diagonal form by the block--diagonal $SU(6)$ matrices $Q_{L,R}$. From these,
one constructs the CKM matrix $V= U^\dag_L D_L$. Carrying out the vectorial
phase changes on the $q_{L,R \ts i}$ required to put $V$ in the standard
Harari--Leurer form with the single CP--violating phase $\delta_{13}$, one
obtains~\cite{harari,pdg}
\bea\label{eq:CKMmat}
V  &\equiv& \left(\ba{ccc}
      V_{ud} & V_{us} & V_{ub}\\
      V_{cd} & V_{cs} & V_{cb}\\
      V_{td} & V_{ts} & V_{tb}\\
      \ea\right)\\ \nn\\
 &=&  \left(\ba{lll}
      c_{12\ts} c_{13} 
      & s_{12\ts} c_{13}
      & s_{13\ts}e^{-i\delta_{13}}\\ 
      -s_{12\ts}c_{23}-c_{12\ts}s_{23\ts}s_{13\ts} e^{i\delta_{13}}
      & c_{12\ts}c_{23}-s_{12\ts}s_{23\ts}s_{13\ts} e^{i\delta_{13}}
      & s_{23\ts}c_{13\ts} \\
      s_{12\ts}s_{23}-c_{12\ts}c_{23\ts}s_{13\ts} e^{i\delta_{13}}
      & -c_{12\ts}s_{23}-s_{12\ts}c_{23\ts}s_{13\ts} e^{i\delta_{13}}
      & c_{23\ts}c_{13\ts}\\
      \ea\right) \ts.\nn
\eea
Here, $s_{ij} = \sin\theta_{ij}$, and the angles $\theta_{12}$,
$\theta_{23}$, $\theta_{13}$ lie in the first quadrant. Additional
CP--violating phases appear in $U_{L,R}$ and $D_{L,R}$ and they are rendered
observable  by ETC and TC2 interactions. We will study their contribution to
$\epsilon$ in Section~6. Before that, we need to discuss the
constraints on $\CM_{u,d}$ and $U_{L,R}$, $D_{L,R}$ imposed by ETC and TC2.

First, limits on flavor--changing neutral current (FCNC) interactions,
especially those mediating $|\Delta S| =2$, require that ETC bosons coupling
to the two light generations have masses $\METC \simge
1000\,\tev$~\cite{etc,tcreview}. These can produce quark masses less than
about $m_s(\METC) \simeq 100\,\mev$ in a walking technicolor theory (see the
Appendix). Extended technicolor bosons as light as 50--$100\,\tev$ are needed
to generate $m_b(\METC) \simeq 3.5\,\gev$. Flavor--changing neutral current
interactions mediated by such light ETC bosons must be suppressed by small
mixing angles between the third and the first two generations.

The most important feature of $\CM_u$ is that the TC2 component of
$\CM_{tt}$, $(m_t)_{TC2} \simeq 160\,\gev$, is much larger than all its other
elements, all of which are generated by ETC exchange. In particular,
off-diagonal elements in the third row and column of $\CM_u$ are expected to
be no larger than the 0.01--1.0~GeV associated with $m_u$ and $m_c$. Thus,
$\CM_u$ is very nearly block--diagonal and, so, $|U_{L,R \ts t u_i}| \cong
|U_{L,R \ts u_i t}| \cong \delta_{t u_i}$.

The matrix $\CM_d$ has a triangular or nearly triangular structure. One
reason for this is the need to suppress $\ol B_d$--$B_d$ mixing induced by
the exchange of ``bottom pions'' of mass $M_{\pi_b} \sim
300\,\gev$~\cite{kominis,bbhk}. Furthermore, since $U_L$ is block--diagonal,
the observed intergenerational mixing in the CKM matrix must come from the
down sector. These requirements are met when the $d_R,s_R \leftrightarrow
b_L$ elements of $\CM_d$ are much smaller than the $d_L,s_L \leftrightarrow
b_R$ elements. In Ref.~\cite{tctwoklee}, the strong topcolor $U(1)$ charges
were chosen to exclude ETC interactions that induce $\CM_{db}$ and
$\CM_{sb}$. This makes $D_R$, like $U_{L,R}$, nearly $2\times 2$ times
$1\times 1$ block--diagonal.

From these considerations and $V_{tb} \cong 1$, we have
\be\label{eq:ckm}
 V_{td_i} \cong V^*_{tb} \ts V_{td_i} \cong U_{L tt} D^*_{L bb}
  U^*_{L tt}  D_{L bd_i} \cong D^*_{L bb} D_{L bd_i} \ts.
\ee
This relation, which is good to 10\% (see Section~4.2 for examples), will be
used in Section~6 to put strong limits on the TC2 $V_8$ and $Z'$ masses from
$\ol B_d$--$B_d$ mixing~\cite{blt}.

One more interesting property of the quark alignment matrices is this: The
vacuum energy $E_{Tq}$ is minimized when the elements of $U$ and $D$ have
almost the same rational phases as $\CM_u$ and $\CM_d$ do. In particular, all
the large diagonal elements of $U, D$ have rational phases (see
Section~4.2). This is generally not true of $U_{L,R}$ and $D_{L,R}$
individually. However, since $Q_{ii} = \sum_j Q_{L ij} Q^*_{R ij}$ ($Q =
U,D$) has a rational phase, $E_{Tq}$ is likely to be minimized when each term
in the sum has the same rational phase. Thus, like DNA, in which the patterns
of the two strands are linked,
\be\label{eq:dna}
\arg Q_{L ij} - \arg Q_{L ik} = \arg Q_{R ij} - \arg Q_{R ik}
\quad ({\rm mod} \pi)
\quad {\rm for} \,\,\,i,j,k = u,c,t \,\,\, {\rm or} \,\,\,d,s,b \ts.
\ee
In particular, $\arg V_{td_i} \cong \arg D_{L bd_i} - \arg D_{L bb} \cong
\arg D_{R bd_i} - \arg D_{R bb}$ (mod $\pi$) for $d_i = d,s,b$.

\subsection*{4.2 Examples}

Our proposal for solving the strong CP problem in technicolor theories rests
on the fact that phases in the technifermion alignment matrices $W =
(W_U,W_D)$ can be different rational multiples of $\pi$, and on the
conjecture that these phases may be mapped by ETC onto the primordial mass
matrix $(\CM_q)_{ij} = \Lambda^{Tq}_{IijJ} \ts W^\dag_{JI} \Delta_T$ so that
$\ol \theta_q = \arg\det(\CM_q) = 0$. Corrections to $\ol \theta_q$ are
expected to be at most $\CO(10^{-10})$. In this section we present two
examples of quark mass matrices for which we have engineered $\ol\theta_q
=0$. They lead to similar alignment and CKM matrices, except that one example
has $\delta_{13} =0$. Nevertheless, as we see in Section~6, both sections
lead to successful calculations of CP--violating parameter $\epsilon$.

\bigskip

\nin {\underbar{\it Model 1:}}

In this model, $\delta_{13} =0$, but CP violation will arise from phases in
$U_{L,R}$ and $D_{L,R}$. The primordial quark mass matrices renormalized at
$\METC$ are taken to be of seesaw form with phases that are multiples of
$\pi/3$:
\bea\label{eq:CMmodela}
\CM_u &=&\left(\ba{lll}        (0,\ts 0) & (200.,\ts 1/3)& (0,\ts 0)\\
                               (15.6,\ts -1/3) & (900,\ts 1) & (0,\ts 0)\\
                               (0,\ts 0) & (0,\ts 0) & (162620,\ts 0)\ea\right)
                               \nn\\ \\
\CM_d &=&\left(\ba{lll}        (0,\ts 0) & (23.3,\ts 0) & (0,\ts 0)\\
                               (21.7,\ts 0) & (102,\ts 1/3) & (0,\ts 0)\\
                               (17.0,\ts 1/3) & (144,\ts 2/3) & (3505,\ts
                               0)\ea\right)
                               \nn \ts.
\eea
The notation is $(|(\CM_q)_{ij}|, \ts \arg [(\CM_q)_{ij}]/\pi)$. Here, we
have made $\arg\det(\CM_u) = \arg\det(\CM_d) = \pi$. We imposed the same kind
of structure on $\CM_u$ as $\ol B_d$--$B_d$ mixing requires of $\CM_d$. The
quark mass eigenvalues may be extracted from $\CM_q$. Their values at $\METC
\sim 10^3\,\tev$ are (in MeV):
\bea\label{eq:mqmodela}
m_u &=& 3.35\ts, \ts\ts\ts m_c = 924\ts, \ts\ts\ts m_t = 162620 \nn\\
m_d &=& 4.74\ts, \ts\ts\ts m_s = 106\ts, \ts\ts\ts m_b = 3508
\eea

The alignment matrices $U = U^\dag_L U_R$ and $D = D^\dag_L D_R$ obtained by
minimizing $E_{Tq}$ are
\bea\label{eq:Qmodela}
U &=&\left(\ba{lll}     (0.973,\ts 0) & (0.232,\ts 1/3) & (0,\ts 0)\\
                        (0.232,\ts -1/3) & (0.973,\ts 1)& (0,\ts 0)\\
                        (0,\ts 0) & (0,\ts 0) & (1,\ts 0)\ea\right)
                        \nn\\ \\
D &=&\left(\ba{lll}     (0.915,\ts -2/3) & (0.404,\ts 0) & (0.0046,\ts -1/3)\\
                        (0.404,\ts 0) & (0.914,\ts -1/3)& (0.0400,\ts -2/3)\\
                        (0.0119,\ts -1/3) & (0.0384,\ts -2/3) & (0.999,\ts
                        0)\ea\right)
                        \nn \ts.
\eea
The cloning of the $\CM_{u,d}$ phases onto $U,D$ is apparent. Diagonalizing
the aligned quark mass matrices yields $Q_{L,R}$:
\bea\label{eq:QLRmodela}
U_L &=&\left(\ba{lll}   (0.9999,\ts -0.859) & (0.0164,\ts 0.141) & (0,\ts 0)\\
                        (0.164,\ts -1.193) & (0.9999,\ts -1.193) & (0,\ts 0)\\
                        (0,\ts 0) & (0,\ts 0) & (1,\ts -0.526)\ea\right) 
                        \nn\\ \nn \\
D_L &=&\left(\ba{lll}   (0.980,\ts 1.141) & (0.199,\ts 1.141) & (0.00485,\ts
                          1.141)\\ 
                        (0.199,\ts -0.192) & (0.979,\ts 0.808) & (0.0412,\ts
                        0.808)\\ 
                        (0.00344,\ts -0.526) & (0.0413,\ts 0.474) &
                        (0.999,\ts -0.526) \ea\right)
                        \nn\\ \\
U_R &=&\left(\ba{lll}   (0.976,\ts -0.859) & (0.216,\ts -0.859) & (0,\ts 0)\\
                        (0.216,\ts -1.193) & (0.976,\ts -0.192) & (0,\ts 0)\\
                        (0,\ts 0) & (0,\ts 0) & (1,\ts -0.526)\ea\right)
                        \nn\\ \nn\\ 
D_R &=&\left(\ba{lll}   (0.977,\ts -0.192) & (0.214,\ts 0.808) &(0.000273,\ts
                          0.808)\\
                        (0.214,\ts 1.141) & (0.977,\ts 1.141) &(0.00122,\ts
                        1.141)\\  
                     (5\times 10^{-6},\ts -0.526) & (0.00125,\ts 0.474)
                     &(1,\ts -0.526) \ea\right) \ts.\nn
\eea
As required, all the mixing in $U_{L,R}$ and $D_R$ is between the first two
generations; mixing of these two with the third generation comes entirely
from $D_L$. A perusal of the phases will reveal differences which are
multiples of $\pi/3$. Finally, the CKM matrix is
\be\label{eq:CKMmodela}
V = \left(\ba{lll}   (0.977,\ts 0) & (0.215,\ts 0) & (0.00552,\ts 0)\\
                        (0.215,\ts 1) & (0.976,\ts 0) & (0.0411,\ts 0)\\
                        (0.00344,\ts 0) & (0.0413,\ts 1) & (0.999,\ts 0)
                        \ea\right) \ts.
\ee
Note its similarity to $D_L$ (including phase differences). This corresponds
to the angles
\be\label{eq:CKMangsmodela}
\theta_{12} = 0.217 \ts, \ts\ts\ts \theta_{23} = 0.0411 \ts, \ts\ts\ts
\theta_{13} = 0.00552 \ts, \ts\ts\ts \delta_{13} = 0\ts.
\ee
The angles $\theta_{ij}$ are in good agreement with those in the Particle
Data Group's book~\cite{pdg}. We will see in Section~6 that, even though
$\delta_{13} = 0$, the CP--violating angles in $D_{L,R}$ can easily account
for the measured value of $\epsilon$.

\bigskip

\nin {\underbar{\it Model 2:}}

The second model is based on a $W$--matrix whose phases are multiples of
$\pi/5$. The primordial quark mass matrices renormalized at $\METC$ are again
taken to be of seesaw form, but we allow off--diagonal terms $|\CM_{ij}| \sim
\sqrt{(|\CM_{ii} \CM_{jj}|)}$ (all masses refer to the ETC contribution only):
\bea\label{eq:CMmodelb}
\CM_u &=&\left(\ba{lll}        (7,\ts 0.2) & (2,\ts -0.4)& (0,\ts 0)\\
                               (100,\ts 0.4) & (890,\ts -0.2) & (0,\ts 0)\\
                               (50,\ts -0.4) & (500,\ts 0.2) & (160000,\ts
                               0)\ea\right)
                               \nn\\ \\
\CM_d &=&\left(\ba{lll}        (8,\ts 0) & (1,\ts -0.2) & (0,\ts 0)\\
                               (25,\ts -0.2) & (100,\ts -0.4) & (0,\ts 0)\\
                               (10,\ts 0) & (140,\ts -0.4) & (3500,\ts
                               0.4)\ea\right)
                               \nn \ts.
\eea
Here, we have made $\arg\det(\CM_u) = \arg\det(\CM_d) = 0$. We again imposed
the same kind of structure on $\CM_u$ as $\ol B_d$--$B_d$ mixing requires of
$\CM_d$. The quark mass eigenvalues are (in MeV):
\bea\label{eq:mqmodelb}
m_u &=& 6.84\ts, \ts\ts\ts m_c = 896\ts, \ts\ts\ts m_t = 160000\nn\\
m_d &=& 7.52\ts, \ts\ts\ts m_s = 103\ts, \ts\ts\ts m_b = 3503
\eea

The alignment matrices $U = U^\dag_L U_R$ and $D = D^\dag_L D_R$ obtained by
minimizing $E_{Tq}$ are
\bea\label{eq:Qmodelb}
U &=&\left(\ba{lll}     (0.994,\ts -0.2) & (0.110,\ts -0.4) & (0.00031,\ts
                          0.4)\\ 
                        (0.110,\ts -0.6) & (0.994,\ts 0.2)& (0.00311,\ts
                        -0.2)\\ 
                        (0.00062,\ts 0.505) & (0.00306,\ts -0.6) & (1,\ts
                        0)\ea\right) 
                        \nn\\ \\
D &=&\left(\ba{lll}     (0.976,\ts 0) & (0.217,\ts 0.2) & (0.00265,\ts
                         -0.0178)\\ 
                        (0.217,\ts -0.8) & (0.975,\ts 0.4)& (0.0389,\ts
                        0.4)\\ 
                        (0.00664,\ts -0.679) & (0.0384,\ts 0.603) &
                        (0.999,\ts -0.4)\ea\right)
                        \nn \ts.
\eea
Again, the cloning of the $\CM_{u,d}$ phases onto the large elements of $U,D$
is apparent. The $Q_{L,R}$ are:
\bea\label{eq:QLRmodelb}
U_L &=&\left(\ba{lll}   (0.994,\ts 0.873) & (0.112,\ts 0.336) & (0.00031,\ts
                          0.535)\\ 
                        (0.112,\ts 0.472) & (0.994,\ts 0.936) & (0.00313,\ts
                        -0.0652)\\ 
                        (0.00063,\ts -0.422) & (0.00308,\ts 0.138) &
                        (1,\ts 0.135)\ea\right)
                        \nn\\ \nn \\
D_L &=&\left(\ba{lll}   (0.970,\ts 0.881) & (0.245,\ts 0.727) & (0.00286,\ts
                          0.535)\\ 
                        (0.245,\ts 0.0810) & (0.969,\ts 0.927) & (0.0400,\ts
                        0.936)\\ 
                        (0.00771,\ts 0.213) & (0.0394,\ts 1.131) & (0.999,\ts
                        0.135) \ea\right)
                        \nn\\ \\
U_R &=&\left(\ba{lll}   (1,\ts 1.073) & (0.00198,\ts 0.534) & (0,\ts 0)\\
                        (0.00198,\ts 0.274) & (1,\ts 0.736) & (1.8\times
                        10^{-5},\ts -0.322)\\ 
                        (0,\ts 0) & (1.8\times 10^{-5},\ts 0.192)& (1,\ts
                        0.135)\ea\right)
                        \nn\\ \nn\\ 
D_R &=&\left(\ba{lll}   (1,\ts 0.881) & (0.0284,\ts 0.727) &(1.7\times
                         10^{-5},\ts 0.661)\\ 
                        (0.0284,\ts -0.319) & (1,\ts 0.527) &(0.00116,\ts
                         0.531)\\ 
                     (1.7\times 10^{-5},\ts 0.611) & (0.00116,\ts -0.470)
                         &(1,\ts 0.535) 
                        \ea\right) \ts.\nn
\eea
The CKM matrix is (compare it to $D_L$)
\be\label{eq:CKMmodelb}
V = \left(\ba{lll}   (0.972,\ts 0) & (0.234,\ts 0) & (0.00315,\ts 0.305)\\
                     (0.233,\ts 0.9999) & (0.971,\ts 8.6\times 10^{-6}) &
                     (0.0431,\ts 0)\\ 
                     (0.00867,\ts 0.0930) & (0.0423,\ts 0.995) & (0.999,\ts 0)
                        \ea\right) \ts.
\ee
This corresponds to the angles
\be\label{eq:CKMangsmodelb}
\theta_{12} = 0.236 \ts, \ts\ts\ts \theta_{23} = 0.0431 \ts, \ts\ts\ts
\theta_{13} = 0.00315 \ts, \ts\ts\ts \delta_{13} = -0.957\ts.
\ee
Again, the angles $\theta_{ij}$ are in reasonable agreement with those
in the Particle Data Group's book. In this model, $\delta_{13}$ is large.

\section*{5. ETC and TC2 Four--Fermion Interactions} 

The FCNC effects that concern us arise from four--quark interactions induced
by the exchange of heavy ETC gauge bosons and of TC2 color--octet
``colorons'' $V_8$ and color--singlet $Z'$. Lepton interactions are not
dealt with here.

At low energies and to lowest order in $\alpha_{ETC}$, the ETC interaction
involves products of chiral currents. Still assuming that the ETC gauge
group commutes with electroweak $SU(2)$, it has the form
\bea\label{eq:HETC}
\chetc &=& \Lambda^{LL}_{ijkl} \left(\ol u'_{Li} \gamma^\mu u'_{Lj} + \ol
d^{\ts \prime}_{Li} \gamma^\mu d^{\ts \prime}_{Lj}\right) \left(\ol u'_{Lk}
\gamma^\mu u'_{Ll} + \ol d^{\ts \prime}_{Lk} \gamma^\mu d^{\ts
\prime}_{Ll}\right) \nn\\
& & \ts +
\left(\ol u'_{Li} \gamma^\mu u'_{Lj} + \ol d^{\ts \prime}_{Li} \gamma^\mu 
d^{\ts \prime}_{Lj}\right)
\left(\Lambda^{u,LR}_{ijkl} \ol u'_{Rk}\gamma^\mu u'_{Rl} +
      \Lambda^{d,LR}_{ijkl} \ol d^{\ts \prime}_{Rk}\gamma^\mu
      d^{\ts \prime}_{Rl}\right) \nn \\
& & \ts + \Lambda^{uu,RR}_{ijkl}\ol u'_{Ri}\gamma^\mu u'_{Rj}\ts\ol
u'_{Rk}\gamma^\mu u'_{Rl} + \Lambda^{dd,RR}_{ijkl}\ol d^{\ts
\prime}_{Ri}\gamma^\mu d^{\ts \prime}_{Rj} \ts\ol d^{\ts
\prime}_{Rk}\gamma^\mu d^{\ts \prime}_{Rl} \nn \\
& & \ts + \Lambda^{ud,RR}_{ijkl} \ol u'_{Ri}\gamma^\mu u'_{Rj}\ts
\ol d^{\ts \prime}_{Rk}\gamma^\mu d^{\ts \prime}_{Rl} \ts,
\eea
where primed fields are electroweak eigenstates. The ETC gauge group contains
technicolor, color and topcolor, and flavor as commuting subgroups~\cite{etc}.
It follows that the flavor currents in $\chetc$ are color and topcolor
singlets. The $\Lambda$'s in $\chetc$ are of order $\getc^2/\METC^2$, whose
magnitude is discussed below, and the operators are renormalized at
$\METC$. Hermiticity of $\chetc$ implies that $\Lambda_{ijkl} =
\Lambda^*_{jilk}$. We assume that this primordial ETC interaction conserves
CP, i.e., that all the $\Lambda$'s are real. When written in terms of mass
eigenstate fields $q_{L,R\ts i} = \sum_j (Q^\dag_{L,R})_{ij} q'_{L,R\ts j}$
with $Q = U,D$, an individual four--quark term in $\chetc$ has the form
\be\label{Hqterm}
  \left(\sum_{i'j'k'l'} \Lambda^{q_1 q_2 \lambda_1 \lambda_2}_{i'j'k'l'}
  \ts\ts   Q^\dag_{\lambda_1\ts ii'} Q_{\lambda_1\ts j'j} \ts
  Q^\dag_{\lambda_2\ts kk'} Q_{\lambda_2\ts l'l}\right) \ts 
  \ol q_{\lambda_1 i} \ts \gamma^\mu \ts q_{\lambda_1 j} \ts\ts
  \ol q_{\lambda_2 k} \ts \gamma_\mu \ts q_{\lambda_2 l} \ts. 
\ee

A reasonable and time--honored guess for the magnitude of the
$\Lambda_{ijkl}$ is that they are comparable to the ETC masses that generate
the quark mass matrix $\CM_q$. We elevate this to a rule: The ETC scale
$\METC/\getc$ in a term involving weak eigenstates of the form $\ol q^{\ts
\prime}_i q'_j \ol q^{\ts \prime}_j q'_i$ or $\ol q^{\ts \prime}_i q'_i \ol
q^{\ts \prime}_j q'_j$ (for $q'_i = u'_i$ or $d^{\ts \prime}_i$) is
approximately the same as the scale that generates the $\ol q^{\ts
\prime}_{Ri} q'_{Lj}$ mass term, $(\CM_q)_{ij}$. A plausible, but
approximate, scheme for correlating a quark mass $m_q(\METC)$ with
$\METC/\getc$ is presented in the Appendix. The results are shown in Fig.~1.
There, $\kappa > 1$ parameterizes the departure from the strict walking
technicolor limit; i.e., $\atc =$ constant and the anomalous dimension
$\gamma_m$ of $\ol T T$ equals one up to the highest ETC mass scale divided
by $\kappa$, and $\gamma_m = 0$ beyond that. The ETC masses run from
$\METC/\getc = 46\,\tev$ for $m_q = 5\,\gev$ to $2.34/\kappa\times
10^4\,\tev$ for $m_q = 10\,\mev$. We rely on Fig.~1 for estimating the
$\Lambda$'s in $\chetc$.

\begin{figure}[t]
 \vspace{6.0cm}
\includegraphics{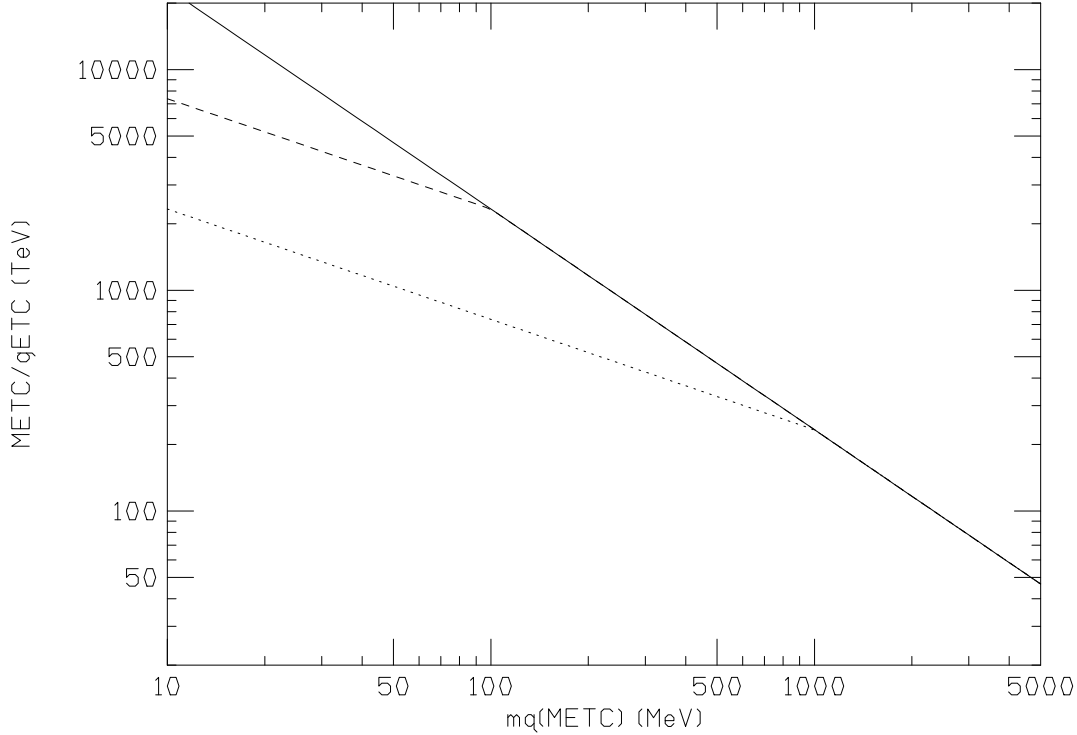}
\vskip 4.0truecm
 \caption{\it
   Extended technicolor scale $\METC/\getc$ as a function of quark mass
   $m_q$ renormalized at $\METC$ for $\kappa = 1$ (solid curve), $\sqrt{10}$
   (dashed), and $10$ (solid); see the Appendix for details.
    \label{fig1} }
\end{figure}

Extended technicolor masses, $\METC/\getc \simge 1000\,\tev$, are necessary,
but not sufficient, to suppress FCNC interactions of light quarks to an
acceptable level. This is especially true for $\Delta M{_K^0}$ and
$\epsilon$~\cite{etc,tcreview}. Thus, we assume that $\chetc$ is {\it
electroweak generation conserving}, i.e.,
\be\label{eq:gencon}
\Lambda^{q_1q_2\lambda_1\lambda_2}_{ijkl} = \delta_{il}\delta_{jk}
\Lambda^{q_1q_2\lambda_1\lambda_2}_{ij} +  \delta_{ij}\delta_{kl}
\Lambda^{\prime \ts q_1q_2\lambda_1\lambda_2}_{ik}\ts.
\ee
Considerable FCNC suppression then comes from off--diagonal elements in the
alignment matrices $Q_{L,R}$.

In all TC2 models, color $SU(3)_C$ and weak hypercharge $\uy$ arise from the
breakdown of the topcolor groups $\suone\otimes\sutwo$ and
$\uone\otimes\utwo$ to their diagonal subgroups. Here, $\suone$ and $\uone$
are strongly coupled, $\sutwo$ and  $\utwo$ are weakly coupled, with the
color and weak hypercharge couplings given by
\bea\label{eq:tccouplings}
g_C &=& {g_1 g_2 \over{\sqrt{g_1^2 + g_2^2}}} \equiv {g_1 g_2\over{g_{V_8}}}
    \equiv g_2 \cos\thc \simeq g_2 \ts; \quad \nn\\
g_Y &=& {g'_1 g'_2 \over{\sqrt{g_1^{\prime \ts 2} + g_2^{\prime \ts 2}}}}
\equiv {g'_1 g'_2\over{g_{Y}}} \equiv g'_2 \cos\thy \simeq g'_2 \ts.
\eea
Top and bottom quarks are $\suone$ triplets. The broken topcolor interactions
are mediated by a color octet of colorons, $V_8$, and a color singlet $Z'$
boson, respectively. By virtue of the different $\uone$ couplings of $t_R$
and $b_R$, exchange of $V_8$ and $Z'$ between third generation quarks
generates a large contribution, $(m_t)_{TC2} \simeq 160\,\gev$, to the
top mass, but none to the bottom mass.

There are two variants of TC2: The ``standard'' version~\cite{tctwohill}, in
which only the third generation quarks are $\suone$ triplets, and the
``flavor--universal'' version~\cite{ccs}, in which all quarks are $\suone$
triplets. In standard TC2, $V_8$ and $Z'$ exchange gives rise to FCNC that
mediate $|\Delta S| =2$ and $|\Delta B|= 2$. In flavor--universal TC2, only
$Z'$ exchange generates such FCNC. We shall write the four--quark interaction
for standard TC2, but our results apply to $Z'$ exchange interactions in
flavor--universal TC2 as well.

The TC2 interaction at energies well below $\Mv$ and $\Mzp$ is
\be\label{eq:HTCT}
\chtct = {g^2_{V_8} \over{2 M^2_{V_8}}} \sum_{A=1}^8 J^{A\mu} J^A_\mu + 
 {g^2_{Z'} \over{2 M^2_{Z'}}} J_{Z'}^\mu J_{Z'\mu} \ts.
\ee
The coloron and $Z'$ currents written in terms of electroweak eigenstate
fields are given by (color indices are suppressed)
\bea\label{eq:JTCT}
 J^A_\mu &=& \cos^2\thc \sum_{i=t,b} \ol q'_i \gamma_\mu \ts
 {\lambda_A\over{2}} \ts  q'_i - \sin^2\thc \sum_{i=u,d,c,s} \ol q'_i
 \gamma_\mu\ts  {\lambda_A\over{2}} \ts q'_i \ts; \nn\\
 J_{Z'\mu} &=& \cos^2\thy J_{1\mu} - \sin^2\thy J_{2\mu} \\
     &\equiv& \sum_{\lambda =L,R} \sum_i \left(\cos^2\thy \ts Y_{1\lambda i}
             - \sin^2\thy \ts Y_{2\lambda i} \right)
           \ol q'_{\lambda i} \gamma_\mu q'_{\lambda i} \ts\ts.\nn
\eea
The $\uone$ and $\utwo$ hypercharges satisfy $Y_{1\lambda i} + Y_{2\lambda i}
= Y_{\lambda i} = 1/6,Q_{EM}$ for $\lambda = L,R$. Consistency with $SU(2)$
symmetry requires $Y_{Lt} = Y_{Lb}$, etc. The suppression of light quark FCNC
requires $Y_{1Li} \equiv Y_{1i}$ for $i=u,d,c,s$ and $Y_{1Ru} = Y_{1Rc}$,
$Y_{1Rd} = Y_{1Rs}$. Remaining FCNC are suppressed by small mixing angles.

\section*{6. TC2 Constraints from $B_d$--$\ol B_d$ Mixing}

If topcolor is to provide a natural explanation of $(m_t)_{TC2}$, the $V_8$
and $Z'$ masses ought to be $\CO(1\,\tev)$. In the Nambu--Jona-Lasinio (NJL)
approximation, the degree to which this naturalness criterion is met is
quantified by the ratio~\cite{cdt}
\be\label{eq:tune}
{\alpha(V_8) + \alpha(Z') - (\alpha^*(V_8) + \alpha^*(Z'))\over{
\alpha^*(V_8) + \alpha^*(Z')}} = {\alpha(V_8)\ts r_{V_8} + \alpha(Z')\ts
r_{Z'} \over {\alpha(V_8)(1-r_{V_8}) + \alpha(Z')(1-r_{Z'})}} \ts.
\ee
Here,
\bea\label{eq:rdef}
\alpha(V_8) &=& {4\alpha_{V_8} \cos^4\thc\over{3\pi}}
         \equiv {4\alpha_C \cot^2\thc\over{3\pi}}, \nn\\
\alpha(Z')  &=& {\alpha_{Z'} Y_{t_L} Y_{t_R} \cos^4\thy\over{\pi}} 
         \equiv {\alpha_Y Y_{t_L} Y_{t_R} \cot^2\thy\over{\pi}}\ts;\\
\tan\thc &=& {g_2\over{g_1}} \ts, \quad \tan\thy = {g'_2\over{g'_1}} \ts,
\quad
r_i = {(m^2_t)_{TC2}\over{M^2_i}} \ts \ln \left({M^2_i
\over{(m^2_t)_{TC2}}}\right) \ts, \hskip0.25truein (i = V_8, Z')\ts;\nn
\eea
and $Y_{t_{L,R}}$ are the $\uone$ charges of $t_{L,R}$. The NJL condition on
the critical couplings for top condensation is $\alpha^*(V_8) + \alpha^*(Z')
= 1$. 

In Ref.~\cite{blt} we showed that, for such large couplings, TC2 is tightly
constrained by the magnitude of $\ol B_d$--$B_d$ mixing. The relevant
observable is the $B^0_L$--$B^0_S$ mass difference. Since $\Gamma_{12} \ll
M_{12}$, it is given by $\Delta M_{B_d} = 2|M_{12}|$. (See Ref.~\cite{buras}
for the standard model description of meson mixing and definitions of the
constants and the functions $S_0$ used below.) The dominant contributions to
$M_{12}$ come from $\ol b' b' \ol b' b'$ terms in Eq.~(\ref{eq:HTCT}).
After Fierzing the product of color--octet currents into a product of
singlets, they give
\be\label{eq:DelMbdTCT}
2 (M_{12})_{\rm TC2} =
{4\pi\over{3}}\left[{\alpha_C \cot^2\thc\over{3 M^2_{V_8}}} + 
{\alpha_Y \cot^2\thy (\Delta Y_L)^2 \over{M^2_{Z'}}} \right]
\ts \eta_B M_{B_d} f^2_{B_d} B_{B_d} \ts (D^*_{Lbb} D_{Lbd})^2
\ts. 
\ee
Here, $\Delta Y_L = Y_{b_L} - Y_{d_L} = Y_{b_L} - Y_{s_L}$ is a difference of
strong $\uone$ hypercharges. We take $\alpha_C \cot^2\thc = \alpha_Y (\Delta
Y_L)^2 \cot^2\thy = 3\pi/8$ so that their sum is NJL--critical if $(\Delta
Y_L)^2 \simeq Y_{t_L} Y_{t_R}$. The QCD radiative correction factor for the
LL product of color--singlet currents is $\eta_B = 0.55 \pm 0.01$. We use
$f_{B_d} \sqrt{B_{B_d}} = (200\pm 40)\,\mev$, where $f_{B_d}$ and $B_{B_d}$
are, respectively, the $B_d$--meson decay constant and bag parameter. This
TC2 contribution is to be added to the standard model one,
\be\label{eq:DelMbdSM}
2(M_{12})_{\rm SM} = {G^2_F\over{6\pi^2}} \eta_B M_{B_d} f^2_{B_d} B_{B_d}
M^2_W S_0(x_t) \ts (V^*_{tb} V_{td})^2 \ts,
\ee
where the top--quark loop function $S_0(x_t) \cong 2.46$ for $x_t =
m_t^2(m_t)/M^2_W$ and $m_t(m_t) = 170\,\gev$.

We argued in Section~4 that the quark--mixing factors in these two
contributions are nearly the same in ETC/TC2 models; see Eq.~(\ref{eq:ckm}).
Thus, the two add without ambiguity and can be used to restrict TC2
parameters. So long as ETC and TC2 do not contribute significantly to the
direct CP--violating parameter $\epsilon'$~\footnote{We are studying this
presumption and, tentatively, believe it is correct.}, we can use its
measurement to limit $|V_{td}|$. Using $\Delta M_{B_d} = (3.11\pm 0.11)
\times 10^{-13}\,\gev$~\cite{pdg}, we found the limits
\be\label{eq:Mlimits}
M_{V_8} \simeq M_Z' \simge 5\,\tev\ts.
\ee
From Eq.~(\ref{eq:tune}), these limits imply that the topcolor coupling
$\alpha(V_8) + \alpha(Z')$ must be within less than 1\% of its critical
value. We regard this tuning to be unnaturally fine.~\footnote{Other limits
on $M_{V_8}$ were obtained in Refs.~\cite{others}.} One way to eliminate this
fine--tuning problem is to invoke the ``top seesaw'' mechanism in which the
topcolor interactions operate on a quark whose mass is several TeV, and the
top's mass comes to it by a seesaw mechanism~\cite{seesaw}.

\section*{7. ETC and TC2 Contributions to the CP--Violating Parameter $\epsilon$}

The CP--violating parameter $\epsilon$ is defined by
\be\label{eq:epsdef}
\epsilon \equiv {A(K_L \ra (\pi\pi)_{I=0}) \over {A(K_S \ra (\pi\pi)_{I=0})}}
= {e^{i\pi/4} \ts {\rm Im} M_{12} \over{\sqrt{2}\ts \Delta M_K}},
\ee
where $2 M_K M_{12} = \langle K^0| \CH_{|\Delta S| = 2} |\ol K^0\rangle$ and
we use the phase convention that $A_0 = \langle (\pi\pi)_{I=0}| \CH_{|\Delta
S| = 1} |K^0\rangle$ is real. Experimentally, $\epsilon = (2.271\pm
0.017)\times 10^{-3} \exp{(i\pi/4)}$~\cite{pdg}. The standard model
contribution to $\epsilon$ is
\be\label{eq:epssm}
\epsilon_{SM} = {e^{i\pi/4} \ts G_F^2 M_W^2 f_K^2 \hat B_K M_K
  \over{3\sqrt{2}\pi^2 \Delta M_K}}
\ts {\rm Im}\left[\lambda_c^{*\ts 2} \eta_1 S_0(x_c)
             + \lambda_t^{*\ts 2} \eta_2 S_0(x_t)
             + 2\lambda_c^*\lambda_t^* \eta_3 S_0(x_c,x_t)\right] \ts,
\ee
where $f_K = 112\,\mev$ is the kaon decay constant, $\hat B_K = 0.80 \pm
0.15$ is the kaon bag parameter, $\lambda_{i=c,t} = V_{id} V^*_{is}$,
$S_0(x_t) \simeq 2.46$, $S_0(x_c) \simeq 1.55\times 10^{-4}$, and
$S_0(x_c,x_t) = 1.40\times 10^{-3}$~\cite{buras}.

Despite the large ETC gauge boson masses of several 1000~TeV and the
stringent $\ol B_d$--$B_d$ mixing constraint leading to TC2 gauge masses of at
least 5~TeV, both interactions can contribute significantly to
$\epsilon$. The main ETC contribution comes from $\ol s' s' \ol s' s'$ 
interactions and is given by 
\bea\label{eq:epsetc}
\epsilon_{ETC} &\simeq& {e^{i\pi/4} \ts f_K^2 M_K \hat B_K\over{3\sqrt{2}\ts
    \Delta M_K}}
    \Biggl\{-\left[\left({M_K\over{m_s + m_d}}\right)^2 + {3\over{2}}\right]
      \Lambda^{LR}_{ss} \ts {\rm Im}\left(D_{Lss} D^*_{Lsd} D_{Rss} D^*_{Rsd}
      \right) \nn\\
 &&\hskip1.2in  + 2\left[\Lambda^{LL}_{ss} \ts {\rm Im}\left(D^2_{Lss}
     D^{*\ts 2}_{Lsd}
           \right)
           +\Lambda^{RR}_{ss} \ts {\rm Im}\left(D^2_{Rss} D^{*\ts
               2}_{Rsd}\right)\right]\Biggr\}.
\eea
Note the suppression of $\CO((\theta_{12})^2)$ from mixing angle
factors. This $\ol s' s' \ol s' s'$ contribution as well as the those from
the standard model and TC2 vanish for Model 1. For that model, ${\rm
Im}(M_{12})_{ETC}$ comes from $\ol s' d' \ol d' s'$ terms and has a form
similar to Eq.~(\ref{eq:epsetc}).

The dominant TC2 contribution comes from $\ol b'_L b'_L \ol b'_L b'_L$
interactions; terms involving $b'_R$ are suppressed by the very small
$D_{Rbd}$ and $D_{Rbs}$:
\be\label{eq:epstct}
\epsilon_{TC2} \simeq {e^{i\pi/4} \ts 4 \pi f_K^2 M_K \hat
                       B_K\over{3\sqrt{2}\ts \Delta M_K}}
        \left[{\alpha_C \cot^2\thc \over{M^2_{V_8}}} + 
              {\alpha_Y (\Delta Y_L)^2 \cot^2\thy \over{M^2_{Z'}}}\right]
             \ts {\rm Im}\left(D^2_{Lbs} D^{*\ts 2}_{Lbd}\right)
             \ts.
\ee
The couplings and mixing angles were defined in Eq.~(\ref{eq:rdef}); we again
take $\alpha_C \cot^2\thc = \alpha_Y (\Delta Y_L)^2 \cot^2\thy =
3\pi/8$~\cite{blt}.

{\begin{table}[t]
\caption{Contributions to $\epsilon \ts e^{-i\pi/4} \times
  10^3$ for various ``models'' of $\CM_q$ with $\ol \theta_q = 0$. Unless
  otherwise indicated, all $\Lambda_{ss} = (2000\,\tev)^{-2}$ and $M_{V_8}
  = M_{Z'} = 10\,\tev$. \label{tab:epsil}}
\begin{center}
%
\begin{tabular}{|c|c|c|c|c|c|l|}
\hline
Model& SM & ${\rm (ETC)_{LR}}$ & ${\rm (ETC)_{LL}}$ & ${\rm (ETC)_{RR}}$ &
${\rm (TC2)_{LL}}$ & Comments\\
\hline\hline
1 & 0 & 2.38 & 0 & 0 & 0 &\begin{minipage}{1.40in} Fit for \\
   $\Lambda_{sd} =$~$(4000\,\tev)^{-2}$
\end{minipage}\\
\hline
1' & 2.28 & 9.61 & 0.88 & 1.02 & 8.34 & \begin{minipage}{1.40in} Fit for \\
  $M_{ETC}\ra$~$\infty$,\\ $M_{V_8,Z'} \ra$~$\infty$
\end{minipage}\\
\hline
2 & -1.98 & 9.44 & -7.68 & -0.11 & -4.57 & \begin{minipage}{1.40in}
  $\epsilon_{ETC+TC2}=$~$4.22$~for \\
  $\Lambda_{ss} =$~$(1250\,\tev)^{-2}\ts,$\\ $M_{V_8,Z'} \ra$~$\infty$
\end{minipage}\\
\hline
2' & 1.97 & -6.30 & 7.76 & 0.05 & 4.52 & \begin{minipage}{1.40in}
  Approximate~fit~for \\$M_{ETC}\ra$~$\infty$,\\ $M_{V_8,Z'} \ra$~$\infty$
\end{minipage}\\
\hline
2'' & -2.02 & 31.25 & -7.92 & -1.21 & -4.68 & \begin{minipage}{1.40in}
  $\epsilon_{ETC+TC2}=$~$4.33$~for \\  $\Lambda_{ss}
  =$~$(2500\,\tev)^{-2}\ts,$\\ $M_{V_8,Z'}=$~$6.9\,\tev$
\end{minipage}\\
\hline
3 & 2.18 & -8.94 & -0.97 & -0.80 & 8.20 &   \begin{minipage}{1.40in}
  $\epsilon_{ETC+TC2}=$~$0.10$~for \\$M_{V_8,Z'}=$~$8.7\,\tev$
\end{minipage}\\
\hline\hline
\end{tabular}
\end{center}
%
\end{table}}

The various contributions to $\epsilon$ for several different ``models'' of
the primordial quark mass matrix $\CM_q$ are given in Table~1. Models~1 and~2
are present as are some related ones (e.g., model~2' is similar to model~2,
but the complex conjugate input $\CM_q$ is used; differences apart from signs
are due to computer round--off error). As indicated, typical ETC masses from
Fig.~1 are used for $\Lambda_{ss}$ and $\Lambda_{sd}$ (the latter for model~1
only). Note that model~1 accounts very well for $\epsilon$ from ETC
interactions alone. ETC interactions that lead to models~1' and~2' are ruled
out. In the other models, $\epsilon$ is easily accounted for because large
cancellations occur among the ETC contributions or between ETC and TC2
contributions. These would be disturbing if we had not already seen them in a
standard context: the large cancellations between QCD and electroweak penguin
terms in the calculation of $\epsilon'/\epsilon$~\cite{buras}.

\section*{8. Summary and Conclusions}

We have presented a dynamical picture of CP nonconservation arising from
vacuum alignment in extended technicolor theories. This picture leads
naturally to a mechanism for evading strong CP violation without an axion or
a massless up quark. We derived complex quark mixing matrices from
ETC/TC2--based constraints on the primordial mass matrices $\CM_u$ and
$\CM_d$. These led to very realistic--looking CKM matrices. We categorized
4--quark contact interactions arising from ETC and TC2 and proposed a scheme
for estimating the strengths of these interactions. Putting this together
with the quark mixing matrices, we showed that TC2 couplings must be tuned
to within 1\% of their NJL critical values to avoid conflict with
$B_d$--$\ol B_d$ mixing. We also calculated contributions to the
CP--violating parameter $\epsilon$, obtaining quite good (or powerfully
constraining) results for a variety of ``models'' of $\CM_q$. Future work
will include calculating $\epsilon'/\epsilon$ and $\sin(2\beta)$ in these
models.

\section*{Acknowledgements}

Most of this results discussed were obtained in collaboration with Gustavo
Burdman, Estia Eichten, and Tongu\c c Rador. I also happily acknowledge the
organizers, Giorgio Belletini, Mario Greco, Giorgio Chiarelli, and the
marvelous secretariat for another successful La Thuile conference. This
research was supported in part by the Department of Energy under
Grant~No.~DE--FG02--91ER40676.

\section*{Appendix. ETC Gauge Boson Mass Scales}

To set the ETC mass scales that enter $\chetc$ in Eq.~(\ref{eq:HETC}), we
assume a model containing $N$ identical electroweak doublets of
technifermions. The technipion decay constant (which helps set the
technicolor energy scale) is then $F_T = F_\pi/\sqrt{N}$, where $F_\pi =
246\,\gev$ is the fundamental weak scale. We estimate the ETC masses in
$\chetc$ by the rule stated in Section~5: The ETC scale $\METC/\getc$ in a
term involving weak eigenstates of the form $\ol q^{\ts \prime}_i q'_j \ol
q^{\ts \prime}_j q'_i$ or $\ol q^{\ts \prime}_i q'_i \ol q^{\ts \prime}_j
q'_j$ (for $q'_i = u'_i$ or $d^{\ts \prime}_i$) is approximately the same as
the scale that generates the $\ol q^{\ts \prime}_{Ri} q'_{Lj}$ mass term,
$(\CM_q)_{ij}$.

The ETC gauge boson mass $\METC(q)$ giving rise to a quark mass
$m_q(\METC)$---an element or eigenvalue of $\CM_q$---is defined
by~\cite{etc,tcreview}
\be\label{eq:qmass}
m_q(\METC) \simeq {g^2_{ETC} \over {M^2_{ETC}(q)}} \condetc \ts.
\ee
Here, the quark mass and the technifermion bilinear condensate, $\condetc$,
are renormalized at the scale $\METC(q)$. The condensate is related to the
one renormalized at the technicolor scale $\LTC \simeq F_T$ by the
equation
\be\label{eq:condrenorm}
\condetc = \condtc \ts \exp\left(\int_{\LTC}^{\METC(q)} \ts {d \mu
\over {\mu}} \ts \gamma_m(\mu) \right) \ts.
\ee
Scaling from QCD, we expect
\be\label{eq:ctc}
\condtc \equiv \Delta_T \simeq 4 \pi F^3_T = 4\pi F^3_\pi/N^{3/2} \ts.
\ee
The anomalous dimension $\gamma_m$ of the operator $\ol T T$ is given in
perturbation theory by
\be\label{eq:gmma}
\gamma_m(\mu) = {3 C_2(R) \over {2 \pi}} \atc(\mu) + O(\atc^2) \ts,
\ee
where $C_2(R)$ is the quadratic Casimir of the technifermion $\sutc$
representation $R$. For the fundamental representation of $\sutc$ to which we
assume our technifermions $T$ belong, it is $C_2(\Ntc) = (\Ntc^2-1)/2\Ntc$.
In a walking technicolor theory, however, the coupling $\atc(\mu)$ decreases
very slowly from its critical chiral symmetry breaking value at $\LTC$, and
$\gamma_m(\mu) \simeq 1$ for $\LTC \simle \mu \simle \METC$.

An accurate evaluation of the condensate enhancement integral in
Eq.~(\ref{eq:condrenorm}) requires detailed specification of the technicolor
model and knowledge of the $\beta(\atc)$--function for large
coupling.\footnote{See Ref.~\cite{multiklrm} for an attempt to calculate this
integral in a walking technicolor model.} Lacking this, we estimate the
enhancement by assuming that
\bea\label{eq:gmm}
\gamma_m(\mu) &= \left\{\ba{ll} 1 & {\rm for} \ts\ts\ts\ts \LTC < \mu <
  \CM_{ETC}/\kappa^2 \\
0 & {\rm for} \ts\ts\ts\ts \mu > \CM_{ETC}/\kappa^2
\ea \right.
\eea
Here, $\CM_{ETC}$ is the largest ETC scale, i.e., the one generating the
smallest term in the quark mass matrix for $\kappa =1$. The number $\kappa >
1$ parameterizes the departure from the strict walking limit (i.e., $\gamma_m
= 1$ constant all the way up to $\CM_{ETC}/\kappa^2$). Then, using
Eqs.~(\ref{eq:qmass},\ref{eq:condrenorm}), we obtain
\bea\label{eq:metc}
{\METC(q)\over{\getc}} &= \left\{\ba{ll} {\sqrt{64\pi^3\alpha_{ETC}}\ts
    F^2_\pi\over {N m_q}}  & {\rm if} \ts\ts\ts\ts \METC(q) <
  \CM_{ETC}/\kappa^2 \\ \\
  \sqrt{{4\pi \CM_{ETC} F^2_\pi \over{\kappa^2 N m_q}}} & {\rm if}
  \ts\ts\ts\ts
  \METC(q) > \CM_{ETC}/\kappa^2
\ea \right.
\eea
To evaluate this, we take $\alpha_{ETC} = 3/4$, a moderately strong value as
would be expected in walking technicolor~\cite{multiklrm}, and $N = 10$, a
typical number of doublets in TC2 models with topcolor
breaking~\cite{tctwoklee}. Then, taking the smallest quark mass at the ETC
scale to be $10\,\mev$, we find $\CM_{ETC} = 7.17\times 10^4\,\tev$. The
resulting estimates of $\METC/\getc$ were plotted in Fig.~1 for $\kappa = 1,
\sqrt{10}$, and 10. They run from $\METC/\getc = 46\,\tev$ for $m_q =
10\,\gev$ to $2.34/\kappa\times 10^4\,\tev/$ for $m_q = 10\,\mev$. Very
similar results are obtained for $\alpha_{ETC} = 1/2$ and $N=8$.

\end{document}